\documentclass[twocolumn,floatfix,eqsecnum,aps,prb]{revtex4-2}
\usepackage{graphicx}
\usepackage{amsmath}
\usepackage{amssymb}
\usepackage{bm}
\usepackage{hyperref}

\begin{document}

\title{Bimodal distribution of delay times and splitting of the zero-bias conductance peak\\
in a double-barrier normal-superconductor junction}
\author{C. W. J. Beenakker}
\affiliation{Instituut-Lorentz, Universiteit Leiden, P.O. Box 9506, 2300 RA Leiden, The Netherlands}
\author{V. A. Zakharov}
\affiliation{Instituut-Lorentz, Universiteit Leiden, P.O. Box 9506, 2300 RA Leiden, The Netherlands}

\date{June 2025}

\begin{abstract}
We formulate a scattering theory of the proximity effect in a weakly disordered SININ junction (S = superconductor, I = insulating  barrier, N = normal metal). This allows to relate the conductance and density of states of the junction to the scattering times $\tau$ (eigenvalues of the Wigner-Smith time-delay matrix). The probability density $P(\tau)$ has two peaks, at a short time $\tau_{\rm min}$ and a late time $\tau_{\rm max}$. The density of states at the Fermi level is the geometric mean of the two times. The splitting of the zero-bias conductance peak is given by $\hbar/\tau_{\rm max}$.
\end{abstract}
\maketitle

\section{Introduction}

Tunneling into a superconductor involves a $2e$ charge transfer, resulting in a quadratic dependence of the conductance $G\propto \Gamma^2$ on the single-electron tunnel probability $\Gamma$. It was discovered in the 1990's that the linear $\Gamma$ dependence of a normal metal is recovered at low bias voltages if a resistive element is placed in series with the superconductor \cite{Wee92,Vol93,Mar93,Hek93,Naz94,Naz99,Bee94,Vol94,Mel94}. This quantum interference effect goes by the name of ``reflectionless tunneling'', because it is as if the Andreev reflected hole can tunnel back into the normal metal with unit probability, so that only one power of $\Gamma$ is paid to add $2e$ to the superconductor. The resulting zero-bias conductance peak has been observed in a variety of materials \cite{Kas91,Cha96,Poi97,Tak02,Qui02,Pop12}.

\begin{figure}[tb]
\centerline{\includegraphics[width=0.8\linewidth]{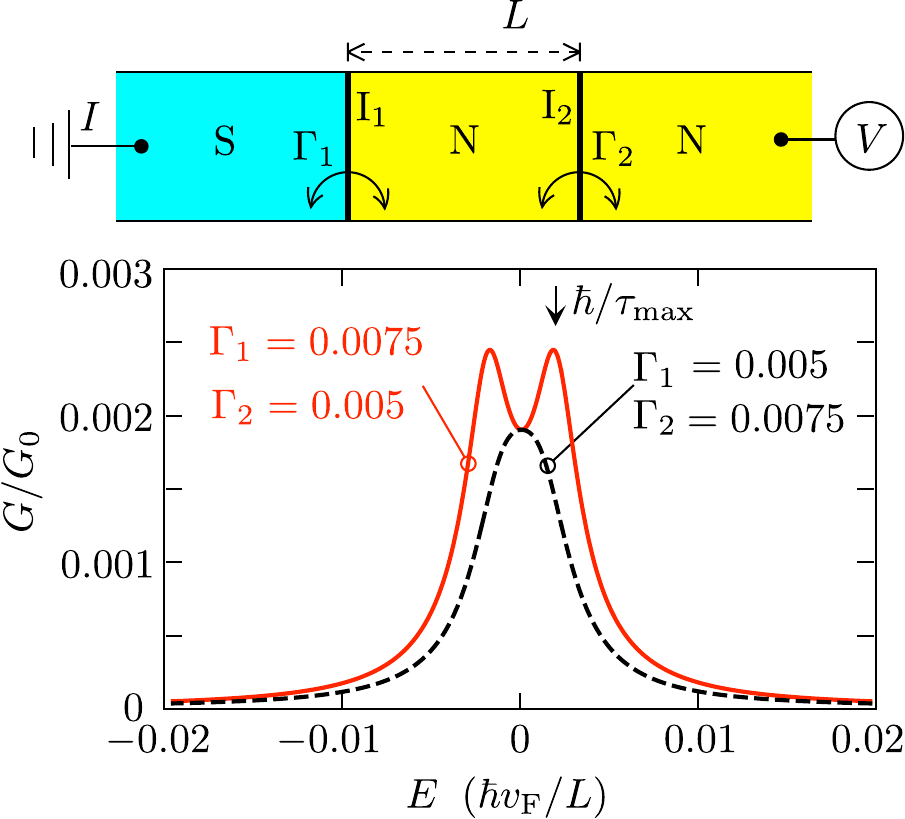}}
\caption{Lower panel: Differential conductance $G=dI/dV$ as a function of voltage bias $V=E/e$ of the $\rm{SI}_1\rm{NI}_2\rm{N}$ junction shown in the upper panel. The curves are computed from Eq.\ \eqref{GEfull} for two combinations of the tunnel probabilities $\Gamma_1,\Gamma_2\ll 1$ through the insulating tunnel barriers. The zero-bias conductance peak from reflectionless tunneling splits when $\Gamma_1/\Gamma_2>\sqrt{2/3}$. The split peaks at $\pm E_{\rm peak}$ are at $\pm\hbar/\tau_{\rm max}$, with $\tau_{\rm max}=(8L/v_{\rm F})\Gamma_2(\Gamma_1^2+\Gamma_2^2)^{-1}$ the late-time peak of the bimodal delay-time distribution.
}
\label{fig_peaks}
\end{figure}

The resistive element that activates reflectionless tunneling can be a disordered normal metal, forming an SIN junction, or it can be a second insulating barrier, forming an $\rm{SI}_1\rm{NI}_2\rm{N}$ junction (S = superconductor, I = insulator, N = normal metal). The $\rm{SI}_1\rm{NI}_2\rm{N}$ junction (see Fig.\ \ref{fig_peaks}) is the superconducting analogue of a Fabry-P\'{e}rot interferometer and serves as a minimal model for the reflectionless tunneling effect \cite{Bee94,Qui02,Les97,Gia01,Big06}. Here we analyze this structure in the weakly disordered regime, where the normal resistance of the junction is dominated by the tunnel barriers (transmission probabilities $\Gamma_1,\Gamma_2$ less than the ratio $\ell/L$ of mean free path and barrier separation). Our objective is to relate the splitting of the zero-bias conductance peak \cite{Cha96,Poi97} that appears when $\Gamma_1\gtrsim\Gamma_2$, to the dynamics of the scattering processes. 

In quantum mechanical scattering theory, the eigenvalues of the Wigner-Smith time delay matrix \cite{Wig55,Smi60}, the socalled proper delay times $\tau$, offer a precise dynamical perspective on the formation of quasibound states and their contribution to observable quantities like conductance and density of states \cite{Car02,Tex16,Lib08}. We find that the delay-time distribution $P(\tau)$ is bimodal, with two distinct peaks: one at a short time $\tau_{\mathrm{min}}\propto\Gamma_2$ and another at a late time $\tau_{\mathrm{max}}\propto \Gamma_2/(\Gamma_1^2+\Gamma_2^2)$. The density of states at the Fermi level is determined by the geometric mean $\sqrt{\tau_{\mathrm{min}} \tau_{\mathrm{max}}}$, while the splitting of the zero-bias conductance peak is governed by the inverse of the late time, $\hbar/\tau_{\mathrm{max}}$. This finding provides a dynamical underpinning of the understanding that the peak splitting results from resonant tunneling through a massively degenerate Andreev level \cite{Big06,Cle00}. 

The scattering theory formulated in Secs.\ \ref{sec_scattering} and \ref{sec_phaseaveraging} is an extension of Ref.\ \onlinecite{Mel94} to include the energy dependence of the scattering matrix $S(E)$, which determines the Wigner-Smith matrix $-i\hbar S^\dagger dS/dE$. Simple closed-form expressions are obtained in Secs.\ \ref{sec_conductance} and \ref{sec_peaksplitting} for the dependence of the differential conductance on the tunnel rates $\Gamma_1,\Gamma_2$, allowing for a precise comparison with the delay-time distribution in Sec.\ \ref{sec_delaytime}. Our calculations are quite elementary, and specific for the SININ geometry, but as we will show in App.\ \ref{comparison} the final results are in complete agreement with more advanced approaches based on Green's functions \cite{Vol93,Naz94,Naz99,Big06} or random-matrix theory \cite{Cle00}.  

\section{Scattering formulation}
\label{sec_scattering}

We begin by summarizing the scattering formulation of the problem \cite{Bee92}. We consider an SININ junction between a normal metal (N) and a superconductor (S), containing a pair of insulating tunnel barriers (${\rm I}_1$, ${\rm I}_2$, see Fig.\ \ref{fig_peaks}). Barrier number 1 is at the interface with the superconductor, barrier number 2 is in the normal-metal region at a distance $L$ from the NS interface. The transmission probability per mode of barrier $i$ is $\Gamma_i\in(0,1)$, taken as mode-independent and energy-independent for simplicity. We take free propagation between the barriers, so we can treat each mode separately.

The single-mode electronic scattering matrix $S(E)$ of the two barriers in series is a $2\times 2$ unitary and symmetric matrix, given by
\begin{widetext}
\begin{align}
S_n(E)={}&\begin{pmatrix}
r'_n(E)&t_n(E)\\
t'_n(E)&r_n(E)
\end{pmatrix}
=\left(1+e^{2iq_n L}\sqrt{1-\Gamma_1}\sqrt{1-\Gamma_2}\right)^{-1}\nonumber\\
{}&\times\begin{pmatrix}
-e^{2iq_nL}\sqrt{1-\Gamma_1}-\sqrt{1-\Gamma_2}&e^{iq_nL}\sqrt{\Gamma_1\Gamma_2}\\
e^{iq_nL}\sqrt{\Gamma_1\Gamma_2}&\sqrt{1-\Gamma_1}+e^{2iq_nL}\sqrt{1-\Gamma_2}
\end{pmatrix}.\label{Sedef}
\end{align}
\end{widetext}
The energy dependence enters via the longitudinal momentum $q_n(E)$ of mode $n$ at energy $E$, measured relative to the Fermi level.

An electron that enters the superconductor at energy $E$ within the superconducting gap $\Delta_0$ is Andreev reflected in the same mode as a hole, with phase shift 
\begin{equation}
\alpha(E)=-\arccos(E/\Delta_0). 
\end{equation}
The hole is scattered by the barriers with scattering matrix $S_n^\ast(-E)$. The scattering sequences combine to produce the probability amplitude $A_n(E)$ for Andreev reflection (electron to hole) by the SININ junction,
\begin{align}
A_n(E)={}&e^{i\alpha(E)}t_n^\ast(-E)\nonumber\\
&\times\left[1-e^{2i\alpha(E)}r_n(E)r_n^\ast(-E)\right]^{-1}t'_n(E).\label{AnEdef}
\end{align}
A similar expression gives the probability amplitude $B_n(E)$ for normal reflection (electron to electron),
\begin{align}
&B_n(E)=r'_n(E)+e^{2i\alpha(E)}t_n(E)\nonumber\\
&\quad\times\left[1-e^{2i\alpha(E)}r_n(E)r_n^\ast(-E)\right]^{-1}r_n^\ast(-E)t'_n(E).\label{BnEdef}
\end{align}
The full reflection matrix is
\begin{equation}
R_n(E)=\begin{pmatrix}
B_n(E)&-A_n^\ast(-E)\\
A_n(E)&B_n^\ast(-E)
\end{pmatrix},\label{RnEdef}
\end{equation}
where we have used electron-hole symmetry to include the amplitude for Andreev reflection from hole to electron and normal reflection from hole to hole.

The differential conductance $G=dI/dV$ at energy $E=eV$ within the gap then follows from a sum over modes $n=1,2,\ldots N$ of the Andreev reflection probability \cite{Blo82},
\begin{equation}
G(E)=\frac{4e^2}{h}\sum_{n=1}^N |A_n(E)|^2.\label{Gdef}
\end{equation}
The conductance quantum $4e^2/h$ contains a factor of two from spin degeneracy and another factor of two because each Andreev reflection transfers $2e$ charge into the superconductor.

\section{Phase averaging}
\label{sec_phaseaveraging}

The energy dependence of the differential conductance is on the scale of the Thouless energy $E_{\rm T}=\hbar/\tau_{\rm dwell}=\hbar v_{\rm F} \Gamma/L$, being the inverse dwell time in the junction region. We assume $L\gg\Gamma\xi_0$, with $\xi_0=\hbar v_{\rm F}/\Delta_0$ the superconducting coherence length. Then $E_{\rm T}\ll\Delta_0$ and for $E\lesssim E_{\rm T}$ we may replace the Andreev reflection phase $\alpha(E)$ by its value at the Fermi level,
\begin{equation}
\alpha(0)=-\pi/2\Rightarrow e^{i\alpha(0)}=-i.
\end{equation}

Because the Thouless energy is small compared to the Fermi energy $E_{\rm F}$, we can linearize the energy dependence of the longitudinal momentum around the Fermi level,
\begin{equation}
q_n(E)=q_n(0)+E/\hbar v_n+{\cal O}(E^2),
\end{equation}
with $v_n$ the longitudinal velocity in mode $n$. The corresponding phase factor $e^{iq_n(E)L}$ in the scattering matrix \eqref{Sedef} factorizes as
\begin{equation}
e^{iq_n(E)L}=e^{i\phi_n}e^{i\varepsilon_n},\;\;\phi_n=q_n(0)L,\;\;\varepsilon_n=EL/\hbar v_n.
\end{equation}

We assume $k_{\rm F}L\gg 1$, $N\gg 1$ and $E\lesssim E_{\rm T}$. Then $\phi_n\gg 1$ is large and strongly $n$-dependent, while $\varepsilon_n\lesssim \Gamma$ is small and only weakly $n$-dependent. This motivates the replacement of the sum $\sum_n$ in Eq.\ \eqref{Gdef} by an integral over $\phi_n$ at constant $\varepsilon_n\approx EL/\hbar v_{\rm F}$,
\begin{equation}
\sum_{n}f(\phi_n,\varepsilon_n)\approx\frac{N}{2\pi}\int_0^{2\pi}d\phi\,f(\phi,\varepsilon),\;\;\varepsilon= EL/\hbar v_{\rm F}.\label{phaseaverage}
\end{equation}

\section{Differential conductance}
\label{sec_conductance}

If we substitute Eq.\ \eqref{Sedef} into Eq.\ \eqref{AnEdef} we find that the phase averaging \eqref{phaseaverage} amounts to an integral of the form \cite{note1}
\begin{equation}
\frac{1}{2\pi}\int_0^{2\pi}d\phi\,\frac{1}{|z-\cos\phi|^2}=-\operatorname{sign}(\operatorname{Re}z)
\frac{\operatorname{Im}(z^2-1)^{-1/2}}{\operatorname{Im}z}.\label{int1}
\end{equation}

For $E\lesssim E_{\rm T}\Rightarrow\varepsilon\ll 1$ we thus obtain
\begin{subequations}
\begin{align}
&G(E)=\operatorname{Im}\frac{G_0 T_{1}\sqrt{T_{2}}\,\varepsilon^{-1}}{\sqrt{T_1+T_2-T_1T_2-4(1+T_2)\varepsilon^2-4i\varepsilon\sqrt{T_2}}},\\
& G_0=\frac{2e^2}{h}N,\;\; T_n=\frac{\Gamma_n^2}{(2-\Gamma_n)^2}\in(0,1),\;\; \varepsilon = \frac{EL}{\hbar v_{\rm F}}\ll 1.
\end{align}
\label{GEfull}
\end{subequations}
Eq.\ \eqref{GEfull} gives the differential conductance of the SININ junction in terms of the individual Andreev reflection probabilities $T_n$ of an SIN junction with tunnel probability $\Gamma_n$ \cite{Bee92}. 

In the zero-bias limit $\varepsilon\rightarrow 0$ we recover the result from Ref.\ \onlinecite{Mel94},
\begin{equation}
G(0)=G_0\frac{2 T_1 T_2}{ (T_1+T_2-T_1T_2)^{3/2}},
\end{equation}
symmetric in $T_1,T_2$. For comparable $T_1\simeq T_2\simeq \Gamma^2 $ this is a zero-bias peak of height $\simeq G_0 \Gamma$, a factor $1/\Gamma$ larger than the conductance $\simeq G_0\Gamma^2$ at large voltages. The switch from a quadratic to a linear scaling in the tunnel probability is the reflectionless tunneling effect.

In App.\ \ref{comparison} we compare with results from the literature. We show that Eq.\ \eqref{GEfull} agrees with the Green's function calculation \cite{Vol93,Naz94,Naz99,Big06} for a weakly disordered SININ junction ($\ell\gg \Gamma L$) in the low-energy regime $E\lesssim E_{\rm T}$. For highly resistive tunnel barriers, $\Gamma_1,\Gamma_2\ll 1$, the SININ junction can be expected to be equivalent to an S--quantum-dot--N junction, and in that limit Eq.\ \eqref{GEfull} agrees with a result from random-matrix theory \cite{Cle00}.

\section{Peak splitting}
\label{sec_peaksplitting}

The $x$-dependence of the function
\begin{equation}
f(x)=\frac{1}{x}\operatorname{Im}(1-\mu x^2-i x)^{-1/2}\label{fdef}
\end{equation}
exhibits a bifurcation at the parameter value $\mu=5/12\equiv\mu_{\rm c}$. For $\mu<\mu_{\rm c}$ there is a single peak at $x=0$, for $\mu>\mu_{\rm c}$ the peak splits into two peaks at $\pm x_{\rm peak}$. For $\mu$ just above threshold the peak position $x_{\rm peak}\approx 1$ is only weakly $\mu$-dependent.

Eq.\ \eqref{GEfull} is of the form \eqref{fdef} with
\begin{equation}
\begin{split}
&\mu=\tfrac{1}{4}(1+1/T_2)(T_1+T_2-T_1 T_2),\\
&x=\frac{4\varepsilon\sqrt{T_2}}{T_1+T_2-T_1T_2},
\end{split}\label{muxdef}
\end{equation}
which for $\Gamma_1,\Gamma_2\ll 1$ reduces to
\begin{equation}
\mu\approx\tfrac{1}{4}(1+\Gamma_1^2/\Gamma_2^2),\;\;
x\approx\frac{8\Gamma_2\varepsilon}{\Gamma_1^2+\Gamma_2^2}.\label{muxdefsmallgamma}
\end{equation}

\begin{figure}[tb]
\centerline{\includegraphics[width=0.8\linewidth]{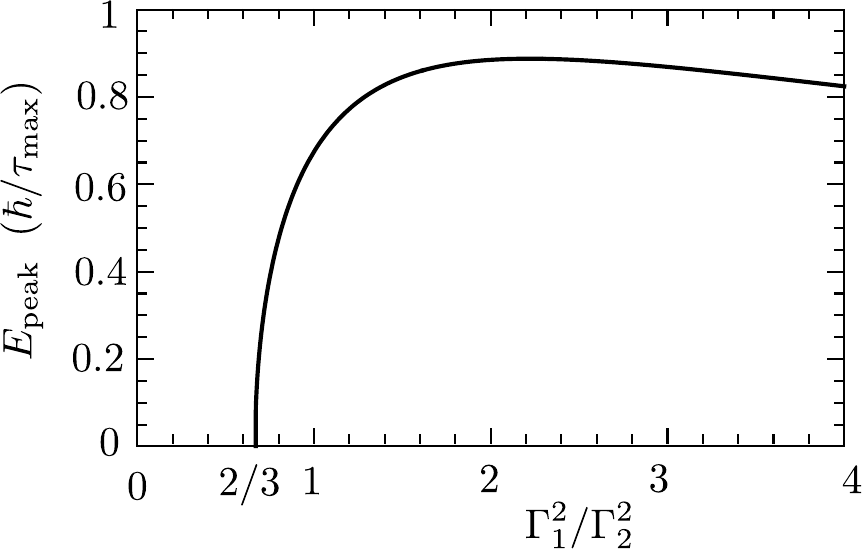}}
\caption{Splitting of the zero-bias conductance peak as a function of the ratio of the tunnel probabilities $\Gamma_1$ (at the NS interface) and $\Gamma_2$ (in the normal region). The curve is obtained by computing the maximum at $\pm E_{\rm peak}$ of $G(E)$ from Eq.\ \eqref{GEfull}, in the small-$\Gamma$ range when $T_1/T_2\approx\Gamma_1^2/\Gamma_2^2\equiv\kappa$. The steep rise at the critical point is the square root $\sqrt{2\kappa-4/3}$. The plot shows the range close to threshold, where $E_{\rm peak}\approx\hbar/\tau_{\rm max}=(\hbar v_{\rm F}/8L)(\Gamma_1^2+\Gamma_2^2)/\Gamma_2$. Far above threshold, for $\kappa\gg 1$, one has $E_{\rm peak}\rightarrow \Gamma_1(\hbar v_{\rm F}/4L)$, independent of $\Gamma_2$.
}
\label{fig_peakenergy}
\end{figure}

We thus find that for $\Gamma_1,\Gamma_2\ll 1$ the zero-bias peak splits when $\Gamma_1/\Gamma_2>\sqrt{2/3}$. Just above this threshold the split peak positions are at $ \pm E_{\rm peak}$, see Fig.\ \ref{fig_peakenergy}, with 
\begin{equation}
E_{\rm peak}\approx \frac{\hbar v_{\rm F}}{L}\,\frac{\Gamma_1^2+\Gamma_2^2}{8\Gamma_2},\;\;\Gamma_2\lesssim\Gamma_1\ll 1.\label{Epeakresult}
\end{equation}
The switch from $\mu<\mu_{\rm c}$ (single peak) to $\mu>\mu_{\rm c}$ (split peak) is visualized in Fig.\ \ref{fig_peaks}.

\section{Delay time distribution}
\label{sec_delaytime}

The delay time in mode $n$ is an eigenvalue of the Wigner-Smith matrix \cite{Wig55,Smi60}, defined in terms of the reflection matrix \eqref{RnEdef} by
\begin{equation}
Q_n(E)=-i\hbar R^\dagger_n(E)\frac{d}{dE}R_n(E).
\end{equation}
We evaluate it at the Fermi level, $E=0$. The Hermitian $2\times 2$ matrix $Q_n(0)$ has two identical eigenvalues,
\begin{equation}
\tau(\phi_n)=\frac{2L}{v_{\rm F}}\frac{\sqrt{T_2}}{1+\sqrt{(1-T_1) (1-T_2)}\cos 2\phi_n}.
\end{equation}

\begin{figure}[tb]
\centerline{\includegraphics[width=0.8\linewidth]{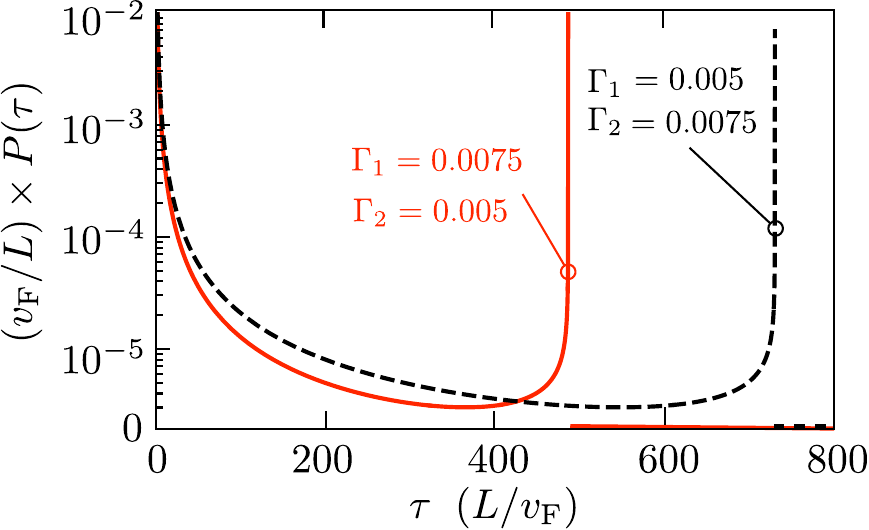}}
\caption{Bimodal delay-time distribution \eqref{Ptauresult} for the two combinations of tunnel probabilities corresponding to the conductance of Fig.\ \ref{fig_peaks}.
}
\label{fig_delay}
\end{figure}

The probability distribution $P(\tau)$ of the delay times follows upon phase averaging,
\begin{subequations}
\begin{align}
P(\tau)={}&\frac{1}{2\pi}\int_0^{2\pi}\delta\bigl(\tau-\tau(\phi)\bigr)\,d\phi\nonumber\\
={}&\frac{1}{\pi \tau}\frac{ \sqrt{\tau_-  \tau_+}}{\sqrt{\tau-\tau_{-}}\sqrt{\tau_{+}-\tau}},\;\;\tau_{-}<\tau<\tau_{+},\\
\tau_{\pm}={}&\frac{(2L/v_{\rm F})\sqrt{T_2}}{T_1+T_2-T_1T_2}\bigl[1\pm\sqrt{(1-T_1)(1-T_2)}\bigr].
\end{align}
\label{Ptauresult}
\end{subequations}
The bimodal distribution is plotted in Fig.\ \ref{fig_delay}, for the tunnel probabilities corresponding to Fig.\ \ref{fig_peaks}. The peak at the shortest and longest delay time $\tau_\pm$ is an inverse square root singularity. This is unusual, for chaotic scattering the delay-time distribution function vanishes at the end points \cite{Bro97,Som01}. One can interpret these two peaks as representing electrons that are reflected quickly by the first barrier, within a time $\tau_-$, or enter the double barrier region and are reflected much later, after a time $\tau_+$.

The average delay time,
\begin{align}
\bar{\tau}={}&\int_{\tau_-}^{\tau_+}\tau P(\tau)\,d\tau\nonumber\\
={}&\sqrt{\tau_- \tau_+}=\frac{2L}{v_{\rm F}}\sqrt{\frac{T_2}{T_1+T_2-T_1T_2}},
\end{align}
equals the density of states per mode at the Fermi level,
\begin{equation}
\rho(0)=\frac{1}{h}\bar{\tau}=\frac{1}{h}\sqrt{\tau_- \tau_+}.\label{rhot}
\end{equation}

For high barriers, $\Gamma_1,\Gamma_2\ll 1$, the shortest and longest delay times are
\begin{equation}
\begin{split}
&\tau_{\rm min}\equiv \tau_-= \tfrac{1}{2}\Gamma_2 L/v_{\rm F},\\
&\tau_{\rm max}\equiv\tau_+=(8L/v_{\rm F})\frac{\Gamma_2}{\Gamma_1^2+\Gamma_2^2}.
\end{split}
\end{equation}
Comparison with Eq.\ \eqref{Epeakresult} shows that the split conductance peak is at an energy
\begin{equation}
E_{\rm peak}\approx \hbar/\tau_{\rm max},\;\;\text{for}\;\;\Gamma_2\lesssim\Gamma_1\ll 1.
\end{equation}

\section{Conclusion}

We have revisited the theory of reflectionless tunneling in a double-barrier normal-superconducting junction, focusing on the splitting of the zero-bias conductance peak. A scattering formulation in a clean junction \cite{Mel94} recovers the key features of the more general Green's function theory \cite{Vol93,Naz94,Naz99,Big06} in the weakly-disordered regime. The merit of the scattering approach is that it readily provides closed-form expressions for the dependence of the peak splitting energy $E_{\rm peak}$ on the tunnel probabilities $\Gamma_1,\Gamma_2$ of the ${\rm SI}_1{\rm NI}_2{\rm N}$ junction. This allowed us to relate the effect to the presence of a peak in the delay-time distribution function at time $\hbar/E_{\rm peak}$.

We have found that the conductance peak splitting is a bifurcation transition at a critical value $\kappa_{\rm c}=2/3$ of the ratio $\kappa=(\Gamma_1/\Gamma_2)^2$. For $\kappa<\kappa_{\rm c}$ the maximum of the conductance is at zero voltage, for $\kappa>\kappa_{\rm c}$ it switches to a nonzero voltage, see Fig.\ \ref{fig_peakenergy}. 

One might wonder whether the density of states $\rho(E)$ in the junction shows any qualitative change at the critical $\kappa$-ratio. The answer is no, see App.\ \ref{app_DOS}. The proximity to the superconductor suppresses the density of states in the normal region by a factor $(1+\kappa)^{-1/2}$, and the whole profile of $\rho(E)$ evolves smoothly as a function of $\kappa$.

All of this is for comparable $\Gamma_1,\Gamma_2$, so $\kappa$ of order unity. The qualitative difference between differential conductance and density of states becomes insignificant for $\kappa\gg  1$, when an excitation gap $\Delta_{\rm min}\simeq \Gamma_1\hbar v_{\rm F}/L$ opens up in the normal region and the conductance peak lines up with the band edge, $E_{\rm peak}\approx\Delta_{\rm min}$, as discussed in Ref.\ \onlinecite{Cle00} for a quantum dot geometry.

\acknowledgments

This work was supported by the Netherlands Organisation for Scientific Research (NWO/OCW), as part of Quantum Limits (project number {\sc summit}.1.1016). Discussions with Anton Akhmerov have been very helpful.

\appendix

\section{Comparison with results in the literature}
\label{comparison}

\subsection{Comparison with the Green's function theory}

The Green's function theory \cite{Vol93,Naz94,Naz99,Big06} gives general expressions for the conductance of a disordered SININ junction. We can compare with our results for a clean junction.

The characteristic energy scales of the problem are the superconducting gap $\Delta_0$, the inverse diffusion time through the normal region, $E_{\rm diffusion}=\hbar v_{\rm F}\ell/L^2$, and the Thouless energy of the tunnel barriers, $E_{\rm T}=\hbar v_{\rm F}\Gamma/L$. Our results apply near zero voltage bias, $|eV|\lesssim E_{\rm T},$ in the regime that the junction is weakly disordered,
\begin{equation}
\begin{split}
&E_{\rm T}\ll E_{\rm diffusion}\ll \Delta_0,\\
&\Gamma\ll \ell/L\ll L/\xi_0. 
\end{split}\label{regime}
\end{equation}

In that regime the Green's function theory gives the following results \cite{Big06}. Define
\begin{widetext}
\begin{subequations}
\begin{align}
&Z_{\rm N}(\theta)=[1+\tfrac{1}{2}\Gamma_2 (\cosh\theta-1)]^{-1},\;\;
Z_{\rm S}(\theta)=[1+\tfrac{1}{2}\Gamma_1 (i\sin\theta-1)]^{-1},\\
&C(\theta)=-\frac{\Gamma_2\tanh(\operatorname{Re}\theta)}{8\varepsilon\sin(\operatorname{Im}\theta)}\bigl[(2-\Gamma_2)\cos(\operatorname{Im}\theta)+\Gamma_2\cosh(\operatorname{Re}\theta)\bigr]|Z_{\rm N}(\theta)|^2,
\end{align}\label{Green1}
\end{subequations}
depending on a complex parameter $\theta$ that solves the transcendental equation
\begin{equation}
\Gamma_2 Z_{\rm N}(\theta)\sinh\theta-4i\varepsilon\sinh\theta+i\Gamma_1 Z_{\rm S}(\theta)\cosh\theta=0.
\end{equation}
The differential conductance of the ${\rm SI}_1{\rm NI}_2{\rm N}$ junction is then given by \cite{Big06}
\begin{equation}
G=\tfrac{1}{2}G_0 C(\theta)\Gamma_2\cosh(\operatorname{Re}\theta)\bigl[\Gamma_1\cosh(\operatorname{Re}\theta)-(2-\Gamma_1)\sin(\operatorname{Im}\theta)\bigr]|Z_{\rm S}(\theta)|^2,\label{Green2}
\end{equation}
at rescaled energy $\varepsilon=E L/\hbar v_{\rm F}$.
\end{widetext}

\begin{figure}[tb]
\centerline{\includegraphics[width=0.8\linewidth]{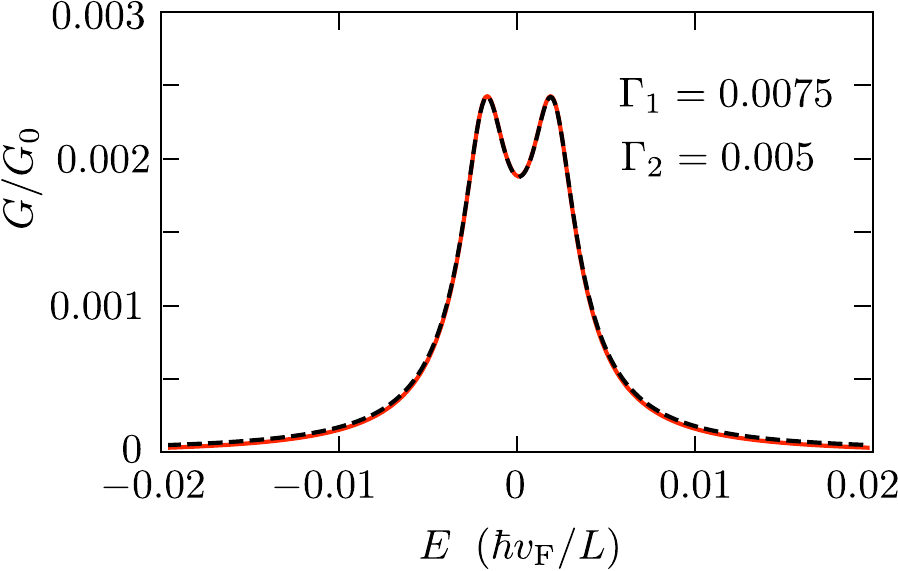}}
\caption{Dashed black curve: Differential conductance of a weakly disordered ${\rm SI}_1{\rm NI}_2{\rm N}$ junction according to the Green's function theory in the regime \eqref{regime}, obtained from Eq.\ \eqref{Green2} after a numerical solution of Eq.\ \eqref{Green1}. The red solid curve is our low-energy approximation \eqref{GEfull} in a clean junction.
}
\label{fig_comparison}
\end{figure}

A numerical solution of these equations agrees very well for $E\lesssim E_{\rm T}$ with our Eq.\ \eqref{GEfull}, see Fig.\ \ref{fig_comparison}. 

While these two theoretical approaches are equivalent in the weakly disordered regime \eqref{regime}, the Green's function theory applies also to stronger disorder. From Refs.\ \onlinecite{Vol93,Mel94,Naz99} we know that the zero-bias conductance peak vanishes if $\ell/L$ becomes much smaller than $\Gamma$. The assumption of mode-independent tunnel probabilities can also be relaxed in the Green's function theory, but this has no qualitative effect on the zero-bias peak. In particular, Ref.\ \onlinecite{Big06} shows that it persists for a disordered tunnel barrier. 

\subsection{Comparison with random-matrix theory}

Clerk, Brouwer, and Ambegaokar \cite{Cle00} have used random-matrix theory to study the proximity effect in a quantum dot connected by a pair of point contacts to a normal and a superconducting reservoir. It is expected that their system corresponds to the weakly disordered ${\rm SI}_1{\rm NI}_2{\rm N}$ junction in the limit $\Gamma_1,\Gamma_2\ll 1$. In that limit Eq.\ \eqref{GEfull} reduces to
\begin{equation}
G(E)=\operatorname{Im}\frac{G_0 \Gamma_1^2\Gamma_2\,\varepsilon^{-1}}{4\sqrt{\Gamma_1^2+\Gamma_2^2-16\varepsilon^2-8i\varepsilon\Gamma_2}},
\end{equation}
After some algebra this can be rewritten as
\begin{subequations}
\begin{align}
&G(E)=\frac{\sqrt{2}G_0 \Gamma_1^2 \Gamma_2^2}{\Xi\sqrt{\Xi+\Gamma_1^2+\Gamma_2^2-16 {\varepsilon}^2}},\\
&\Xi=\sqrt{2 \Gamma_2^2 \left(16 {\varepsilon}^2+\Gamma_1^2\right)+\left(16 {\varepsilon}^2-\Gamma_1^2\right)^2+\Gamma_2^4},
\end{align}
\end{subequations}
in agreement with Ref.\ \onlinecite{Cle00}.

\section{Density of states}
\label{app_DOS}

The density of states $\rho(E)$ in the SININ junction can be determined from the full reflection matrix \eqref{RnEdef},
\begin{equation}
\rho(E)=\frac{1}{2\pi i}\sum_{n}\frac{d}{dE}\ln\det R_n(E).
\end{equation}
Substitution of Eqs.\ \eqref{AnEdef} and \eqref{BnEdef} gives
\begin{widetext}
\begin{equation}
\frac{1}{2\pi i}\frac{d}{dE}\ln\det R_n(E)=\frac{\rho_0\sqrt{T_2}\bigl(1+\sqrt{1-T_1} \sqrt{1-T_2} \cos 2 \varepsilon   \cos 2 \phi_n\bigr) }{1+2 \sqrt{1-T_1} \sqrt{1-T_2} \cos 2 \varepsilon   \cos 2 \phi_n  -(1-T_2) \sin ^2 2 \varepsilon  +(1-T_1) (1-T_2) \cos ^2 2 \phi_n  },
\end{equation}
\end{widetext}
with $\rho_0=2L/\pi\hbar v_{\rm F}$.

\begin{figure}[tb]
\centerline{\includegraphics[width=0.8\linewidth]{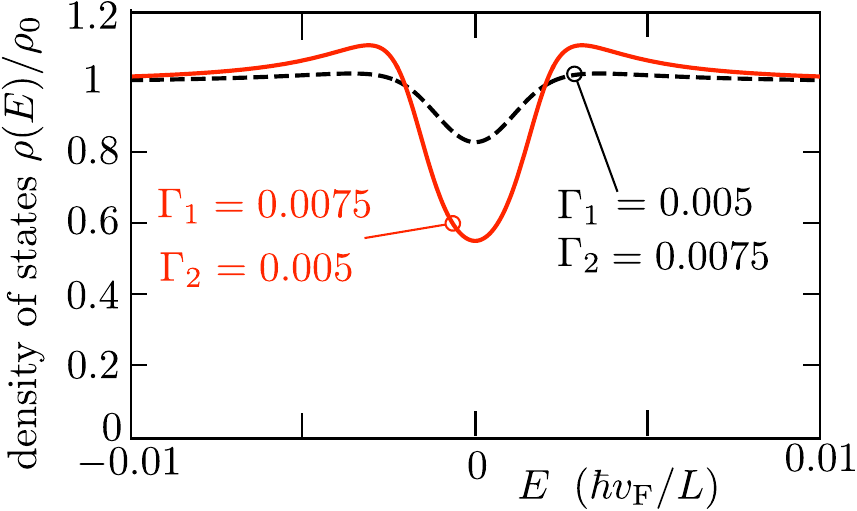}}
\caption{Density of states of the SININ junction, computed from Eq.\ \eqref{rhofull} for the same combination of tunnel probabilities as in Fig.\ \ref{fig_peaks}.
}
\label{fig_DOS}
\end{figure}

The phase average \eqref{phaseaverage} can be carried out with the help of Eq.\ \eqref{int1} and the similar
\begin{equation}
\frac{1}{2\pi}\int_0^{2\pi}d\phi\,\frac{\cos\phi}{|z-\cos\phi|^2}=-\operatorname{sign}(\operatorname{Re}z)
\frac{\operatorname{Im}z(z^2-1)^{-1/2}}{\operatorname{Im}z}.\label{int2}
\end{equation}
The resulting density of states, per mode and for a single spin direction, is
\begin{align}
&\rho(E)=\rho_0\operatorname{Im}\frac{2\varepsilon+i\sqrt{T_2}}{\sqrt{T_1+T_2-T_1T_2-4(1+T_2)\varepsilon^2-4i\varepsilon\sqrt{T_2}}},\label{rhofull}
\end{align}
plotted in Fig.\ \ref{fig_DOS}.

At the Fermi level, $E\rightarrow 0$, one finds
\begin{equation}
\rho(0)=\rho_0\sqrt{\frac{T_2}{T_1+T_2-T_1T_2}}\approx\rho_0(1+\Gamma_1^2/\Gamma_2^2)^{-1/2},
\end{equation}
for $\Gamma_1,\Gamma_2\ll 1$.


\begin{thebibliography}{99}
\bibitem{note1} The integral \eqref{int1} exists for all $z\in\mathbb{C}$ not in the interval $[-1,1]$. For purely real or purely imaginary $z$ the right-hand-side of Eq.\ \eqref{int1} should be understood in the limit of an infinitesimal real and imaginary part, where the branch cut of the square root is taken along the negative real axis.
\end{thebibliography}
\end{document}